\documentclass[journal]{IEEEtran}
\ifCLASSINFOpdf
\else
\fi
\usepackage[numbers]{natbib}
\usepackage{booktabs}
\usepackage[flushleft]{threeparttable}
\usepackage{graphicx}
\usepackage{enumitem}
\usepackage{array,ragged2e}
\usepackage{array}
\ifCLASSOPTIONcompsoc
    \usepackage[caption=false, font=normalsize, labelfont=sf, textfont=sf]{subfig}
\else
\usepackage[caption=false, font=footnotesize]{subfig}
\fi
\newcolumntype{C}[1]{>{\centering\let\newline\\\arraybackslash\hspace{0pt}}m{#1}}
\usepackage{amssymb}
\usepackage{pifont}
\newcommand{\cmark}{\ding{51}}%
\newcommand{\xmark}{\ding{55}}%
\usepackage{multirow}
\usepackage{amsthm}
\newtheorem*{remark}{Remark}
\usepackage{hyperref}
\begin{document}
%
\title{A Privacy-Preserving Energy Theft \\Detection Model for Effective Demand-Response Management in Smart Grids}
%
%
%

\author{Arwa~Alromih,
        John~A.~Clark,
        and~Prosanta~Gope,~\IEEEmembership{Senior~Member,~IEEE}
\thanks{Authors are with the Department of Computer Science, University of Sheffield, Regent Court, Sheffield S1 4DP, United Kingdom. (E-mails: \mbox{asmalromih1}, john.clark, p.gope@sheffield.ac.uk).}
\thanks{A. Alromih is also with the Information Systems Department, King Saud University, Riyadh, Saudi Arabia.(E-mail: aalromih@ksu.edu.sa)}
\thanks{Manuscript received XXX XX, XXXX; revised XXX XX, XXXX.}}

%
%

\markboth{Journal of \LaTeX\ Class Files,~Vol.~X, No.~X, XXXX~2023}%
{Alromih \MakeLowercase{\textit{et al.}}: Bare Demo of IEEEtran.cls for IEEE Journals}
%



\maketitle

\begin{abstract}
The detection of energy thefts is vital for the  safety of the whole smart grid system. However, the detection alone is not enough since energy thefts can crucially affect the electricity supply leading to some blackouts. Moreover, privacy is one of the major challenges that must be preserved when dealing with clients' energy data. This is often overlooked in  energy theft detection research as most current detection techniques rely on raw, unencrypted data, which may potentially expose sensitive and personal data. To solve this issue, we present a privacy-preserving energy theft detection technique with effective demand management that employs two layers of privacy protection. We explore a split learning mechanism that trains a detection model in a decentralised fashion without the need to exchange raw data. We also employ a second layer of privacy by the use of a masking scheme to mask clients' outputs in order to prevent inference attacks. A \emph{privacy-enhanced version} of this mechanism also employs an additional layer of privacy protection by training a randomisation layer at the end of the client-side model. This is done to make the output as random as possible without compromising the detection performance. For the energy theft detection part, we design a multi-output machine learning model to identify energy thefts, estimate their volume, and effectively predict future demand. Finally, we use a comprehensive set of experiments to test our proposed scheme. The experimental results show that our scheme achieves high detection accuracy and greatly improves the privacy preservation degree.
\end{abstract}


%

\section{Introduction}
%
%
%
%
\IEEEPARstart{T}{he} concept of smart grids (SG) refers to the modernisation of traditional electricity grids to allow dynamic optimisation of operations and maintain reliable and secure electricity infrastructure. SG uses an advanced metering infrastructure (AMI) that utilises digital information and communication technology (ICT) to allow real-time measurement of power demand to be exchanged between all components. These high-resolution electricity data, provided by modern smart meters, help balance supply and demand better. New smart grids also allowed the integration of renewable energy resources at residential levels, which enabled consumers to produce energy and sell it to the grid. The advancements in smart grid technology and its integration with ICT have brought many vulnerabilities related to malicious behaviour, such as energy thefts.
Energy thefts are one of the major causes of non-technical losses (NTL) during electricity transmission and distribution \cite{liu2020hidden}.
They are defined as any illegal energy use that violates contract terms. This can be achieved through any physical activity such as wiretapping or digitally manipulating meter readings to pay less or get paid more, in case of selling to the grid \cite{yan2021performance}.
Globally, energy thefts are the most cause of financial losses in the energy market, and it has been reported that around \$1 to \$6 billion dollars are lost yearly in the UK and the US combined due to these attacks  \cite{glauner2017challenge}. 

Energy thefts have non-financial consequences. For instance, they were the cause of 15 different blackout incidents in the US in 2017 alone \cite{eaton2017report}. However, the solution is generally sought solely through detection measures. Moreover, energy thefts can severely affect demand management, leading to mis-forecasting of future supply. By underestimating the electricity supply, some grid regions will experience blackouts. Overestimating the supply leads to an often infeasible need for storage, and batteries are expensive. Traditionally, utilities could detect energy thieves by comparing the aggregated electricity consumption and the supplied energy. If there is a mismatch, technicians need to physically visit the suspicious community to check for any physical breaches to the meters. However, the widespread deployment of smart meters and the help of the two-way communication provided by ICT have enabled electricity utilities to monitor consumption and generation in real time. These fine-grained data exchanged between the grid's components have been the primary source for remotely detecting energy thefts. Most of these approaches mainly use machine learning models. 

Although data-driven approaches have proven to be very accurate in detecting energy theft attacks \cite{glauner2017challenge}, the use of high-resolution electricity data poses some serious privacy concerns \cite{gope2018lightweight,glauner2017challenge}. They might reveal valuable information, such as what appliances are operating, whether someone is present in the house, or even a pattern of behaviour. Therefore, theft detection mechanisms must consider the customers' privacy.

\subsection{Related Work and Motivation}\label{related_work}
Over recent decades, several works have been proposed in the literature on demand-response management in smart grids, such as \cite{bahrami2020deep,xiao2021dynamic,akyol2022avoiding}. Authors in \cite{gope2018lightweight,hassan2022differentially} had also considered privacy-preserving approaches for their solutions. However, all of these solutions assumed that the data supplied by the clients were genuine. None has considered the case when energy thefts are present in the system and to what degree this theft would impact demand-supply management.
Similarly, most proposed energy theft detection (ETD) techniques do not consider privacy and use energy usage data in its raw form. However, since such use of data can lead to privacy risks, there was a need to address privacy  in energy theft detection research. A few papers have addressed privacy-preserving energy theft detection in SG. These can be viewed under two broad categories: the first use cryptographic-based methods, while the others are based on privacy-preserving machine learning (PPML) techniques. We now elaborate on the strengths and weaknesses of both approaches.

The first work to study the privacy issue in energy theft detection was introduced by \citet{salinas2013privacy} in \citeyear{salinas2013privacy}. The authors designed three distributed privacy-preserving approaches to identify fraudulent users based on two well-known decomposition algorithms: LU and QR factorization. These algorithms can solve a linear system of equations that corresponds to the energy consumption data of consumers, which must match the total load consumption measured by the collector at each time interval. 
Although this was the first work to look at privacy in ETD, the work did not consider the issue of technical losses.
Following their work, \citet{salinas2015privacy} have also proposed another privacy-preserving energy theft detection based on state estimation. State estimation is a technique that uses a mathematical algorithm known as the Kalman filter to estimate the system's current state at various points where energy thefts are detected by comparing the estimated state variables with the actual measurements. In their proposed work, the authors introduce a decomposed version of the Kalman filter that can hide energy measurements and preserves users' privacy. However, this proposed loosely coupled filter can be employed in small areas of microgrids as the complexity would increase as the size of the grid increases. In addition, according to \cite{jokar2015electricity}, the proposed scheme can only detect thefts with consecutive reduction reads (i.e. when the meter readings show a consistent decrease in consumption over a period of time), while in reality, there are different types of energy thefts. 
\citet{wen2018state} investigated using a recursive filter based on state estimation to estimate users' energy consumption and detect any abnormal behaviours. Privacy is achieved by using the Number Theory Research Unit (NTRU) encryption algorithm to encrypt users' data. However, the scheme assumes that aggregators are trusted entities which is not always the case since most aggregators are third-party companies that are not governed by any authority. 

In \cite{richardson2016privacy}, the Paillier crypto-system was used to preserve the privacy of their proposed ETD. Euclidean distance measures between energy readings over a day were used to detect abnormalities and frauds without revealing any valuable information. Another Paillier-based privacy system was introduced by \citet{yao2019energy}. In their security and privacy analysis, the authors state that the proposed detection algorithm achieves confidentiality, integrity, and data privacy by using encryption and digital signing. However, it is known that this is entirely dependent on the encryption mechanism strength.
\citet{nabil2019ppetd} proposed a secure multiparty computation-based energy theft detection to preserve the privacy of energy readings. The scheme uses secret-sharing techniques to allow smart meters to send masked data. The detection of energy thefts is done online, where the smart meter and the system operator need to run a convolutional neural network (CNN) model. Hence, this solution introduces computational and communication overheads. To overcome the need for running the detection model in both parties in parallel, the authors in \cite{ibrahem2020efficient} used a functional encryption algorithm to encrypt energy readings where energy theft detection, billing and load management are all done without revealing the individuals' readings.

\citet{wen2021feddetect} have designed a federated learning-based energy theft detector with multiple local detection stations trained in a federated fashion. The model is then used to detect energy thefts from local users. To preserve the privacy of the local users' data, a local differential privacy algorithm is used to distribute the energy usage data of the grid's users. While this federated approach can preserve privacy, it introduces additional communication and computation complexity. Additionally, the scheme requires installing additional detection stations in the system.
Another federated learning solution was introduced recently in \cite{ashraf2022feddp}, where a novel federated voting classifier, namely ensemble learning, is used. This scheme assumes that the use of federated learning preserves privacy. However, it has been proven that FL, on its own, can not guarantee high levels of privacy and is very vulnerable to poisoning attacks, feature leakage, and reconstruction attacks \cite{li2020federated,xia2021survey,jin2021cafe}.  Recently, a blockchain-based privacy-preserving energy theft detection was proposed in \cite{muzumdar2022designing}. Energy thefts are detected by comparing the aggregated consumption reports with the energy supplied. Users share their energy consumption privately using energy contracts in a ledger.

\begin{table*}[ht]
\begin{center}
\begin{threeparttable}

\caption{Summary of The Related Work}
\label{tab:related_work}
\renewcommand{\arraystretch}{1.75}
\setlength\tabcolsep{0pt} 
\begin{tabular*}{0.92\textwidth}
{@{\extracolsep{\fill}} p{0.16\textwidth} p{0.24\textwidth}@{\hskip 6pt}p{0.22\textwidth}ccccccc}
\toprule
\Centering{\multirow{2}{*}{\textbf{Scheme}}} &
\Centering{\multirow{2}{0.23\textwidth}{\centering\textbf{Energy Theft Detection Approach}}} &
\Centering{\multirow{2}{0.19\textwidth}{\centering\textbf{Privacy-Preserving Approach}}} &
\multicolumn{7}{c}{\textbf{Supported Features}} \\
\cline{4-10}
&&&F1\tnote{*}&F2\tnote{*}&F3\tnote{*}&F4\tnote{*}&F5\tnote{*}&F6\tnote{*}&F7\tnote{*}\\
\midrule
    \citet{salinas2013privacy} & Linear system of equations & Decomposition algorithms &  
    \cmark & \xmark & \xmark & \xmark & - &\xmark&\xmark\\ \hline
    \citet{salinas2015privacy} & State estimation & Decomposed Kalman filter & 
    \xmark & \xmark & \cmark & - & - &\xmark\tnote{**}&\xmark
    \\    \hline

    \citet{wen2018state} & State estimation & NTRU cryptosystem & 
    \cmark & \xmark & \cmark & \xmark & - &\xmark\tnote{**}&\xmark
    \\    \hline
    \citet{richardson2016privacy} & Machine learning (Clustering) & Paillier cryptosystem  & 
 \cmark & \xmark & \xmark & - & - &\xmark&\xmark
    \\    \hline

    \citet{yao2019energy} & Machine learning (CNN\tnote{1}) & Paillier cryptosystem & 
    \cmark  & \xmark & \xmark & \xmark & -&\xmark\tnote{**}&\xmark
    \\    \hline

    \citet{nabil2019ppetd} & Machine Learning (CNN\tnote{1}) & Secure multiparty computation & 
    \cmark & \cmark & \xmark & \xmark & -&\xmark\tnote{**}&\xmark
    \\    \hline

    \citet{ibrahem2020efficient} & Machine-learning (FNN\tnote{2}) & Functional encryption & 
   \cmark & \cmark & \xmark & \cmark & -&\xmark\tnote{**}&\xmark

    \\    \hline

    \citet{wen2021feddetect}& Machine learning (TCN\tnote{3}) & Federated learning and local differential privacy & 
   \cmark & \xmark & \cmark & \xmark & \xmark &\xmark\tnote{**}&\xmark
    \\    \hline
    \citet{ashraf2022feddp}& Machine learning (Ensemble learning) & Federated learning & 
    \cmark & \xmark & \cmark & \xmark & \xmark &\xmark&\xmark

    \\     \hline
    \citet{muzumdar2022designing} & Difference between energy supply and consumption & Bloackchain & 
     \cmark & \xmark & \xmark & \xmark & -&\xmark\tnote{**}&\xmark
    \\
    \hline
    Proposed scheme & Machine learning (Multi-output NN) & Split learning and masking&\cmark&\cmark&\cmark&\cmark&\cmark&\cmark&\cmark\\
    \bottomrule
\end{tabular*}
\smallskip
\scriptsize
\begin{tablenotes}
\item[*]F1: Scalability; F2: Detecting more than one type of energy thefts; F3: Computational efficiency; F4: Communication efficiency; F5: Stronger resilience against inference attacks; F6: Privacy quantitative analysis; and F7: Considering demand-response management after the detection. 
\item[1]CNN: Convolutional Neural Network. \item[2] FNN: Feed-forward Neural Network.
\item[3] TCN: Temporal Convolutional Network.
\item[**] The study only provides qualitative privacy analysis, not a quantitative one.
\end{tablenotes}
\end{threeparttable}
\end{center}
\end{table*}

\textit{Motivation:} Investigating the existing literature on both demand management and energy theft detection research areas reveals two major issues: (i) Although several works have been proposed in demand management for smart grids, they have yet to consider the issue of managing the demand in cases where energy thefts exist. Instead, the existing research has always assumed that all clients are honest and would report reliable readings, which is not always true. 
(ii) On the other hand, although several solutions have been developed for energy theft detection, all the research in this area has focused solely on the detection part and has not gone beyond that. However, it is essential to act upon detecting energy theft, including estimating the amount of stolen energy and considering it while forecasting the future energy demand. In fact, Ofgem\footnote{Ofgem is the Office of Gas and Electricity Markets and it is the energy regulator for Great Britain.} had set some rules for tackling electricity theft, one of which requires all energy suppliers in the UK to \emph{“make accurate estimates of the volume of electricity stolen following detection"} \cite{ofgem2013tackling}. This is an essential post-detection step that no one has considered before, as indicated in Table \ref{tab:related_work}. To the best of our knowledge, we are the first to propose such a step.

Another limitation in the energy theft detection research area is that the proposed schemes that consider customers' privacy lack any quantitative analysis of that privacy, as seen in Table \ref{tab:related_work}. None appear to have addressed whether the privacy protection of an energy theft detector can be quantified. Whenever a privacy-preserving approach is used, such as encryption-based schemes and differential privacy, it is implicitly assumed that it provides strong privacy protection. Also, these solutions require computationally and communicationally expensive operations (e.g. homomorphic encryption), which restricts their suitability for resource-constrained smart meters. On the other hand, differential privacy-based schemes suffer from a privacy-accuracy trade-off. Privacy gained by these schemes is often proven using strict mathematical proofs, whereas it has been shown that they are vulnerable to many privacy attacks \cite{bernau2021comparing,zhao2019differential}. Moreover, the works that aim to achieve privacy using federated learning suffer from the following weaknesses:
\begin{itemize}
    \item Communication is a critical bottleneck in the federated learning architecture. This is because it involves communicating a large amount of data between clients and the server in every round, where every message contains the complete model parameters.
    \item Each client in the federated learning approach needs to have  a large storage capacity and high computational and communication capabilities to run a complete model. However, this is not the case in the smart meter environment, as devices will generally be resource-limited.
    \item Finally, privacy is often a major concern in federated learning. Sharing model updates (i.e. gradient information) instead of the raw data can reveal sensitive information to an eavesdropper or the system entities. This is a major problem faced in federated learning approaches, giving rise to what are typically referred to as “inference attacks". 
\end{itemize}

\noindent As discussed above, none of the existing research in energy theft detection achieves all the important features listed in Table \ref{tab:related_work}. 


\subsection{Our Contribution}
To address all the limitations mentioned above, in this article, we propose a privacy-preserving energy theft detection model that can effectively predict energy demand. Our proposed model not only detects energy thefts of different types but also helps to reliably manage the power demand even in the event of thefts. This is the first work to develop a solution that bridges the gap between the energy theft detection and demand management research areas.

In this article, we first propose an energy theft detection system that preserves users' privacy by using split learning as the architecture of our machine learning approach. This approach detects energy thefts, estimates the amount of stolen energy and manages the future demand even in cases of thefts. Moreover, to avoid the issue of feature leakage, we utilise a lightweight masking approach to lower the chances of any inference attacks. An enhanced privacy-preserving approach is also proposed by using an added neural network layer that is trained to randomise the outputs of the client's part of the model. In both proposed designs, users' data are kept private while the system can still detect energy thefts with high accuracy.

The remainder of this paper is organised as follows.
In Section II, we explain the details of the system and threat models employed in this paper. We further introduce background on split learning, three-tier split learning and distance correlation. In Section III, we give a detailed description of the proposed privacy-preserving energy theft detector with the demand-management model. In Section IV, we show the analytical results and simulations where we analyse both the detection abilities of the proposed approach and the privacy aspect. Finally, we conclude the paper in Section V.

\section{Preliminaries}
In this section, we first describe the system model of the proposed scheme and explain the threat model considered. We then briefly describe the adopted three-tier split learning architecture, followed by a definition of distance correlation and its usage. 


\subsection{System Model}
\begin{figure}[tb]
    \centering
    \includegraphics[width=0.9\columnwidth]{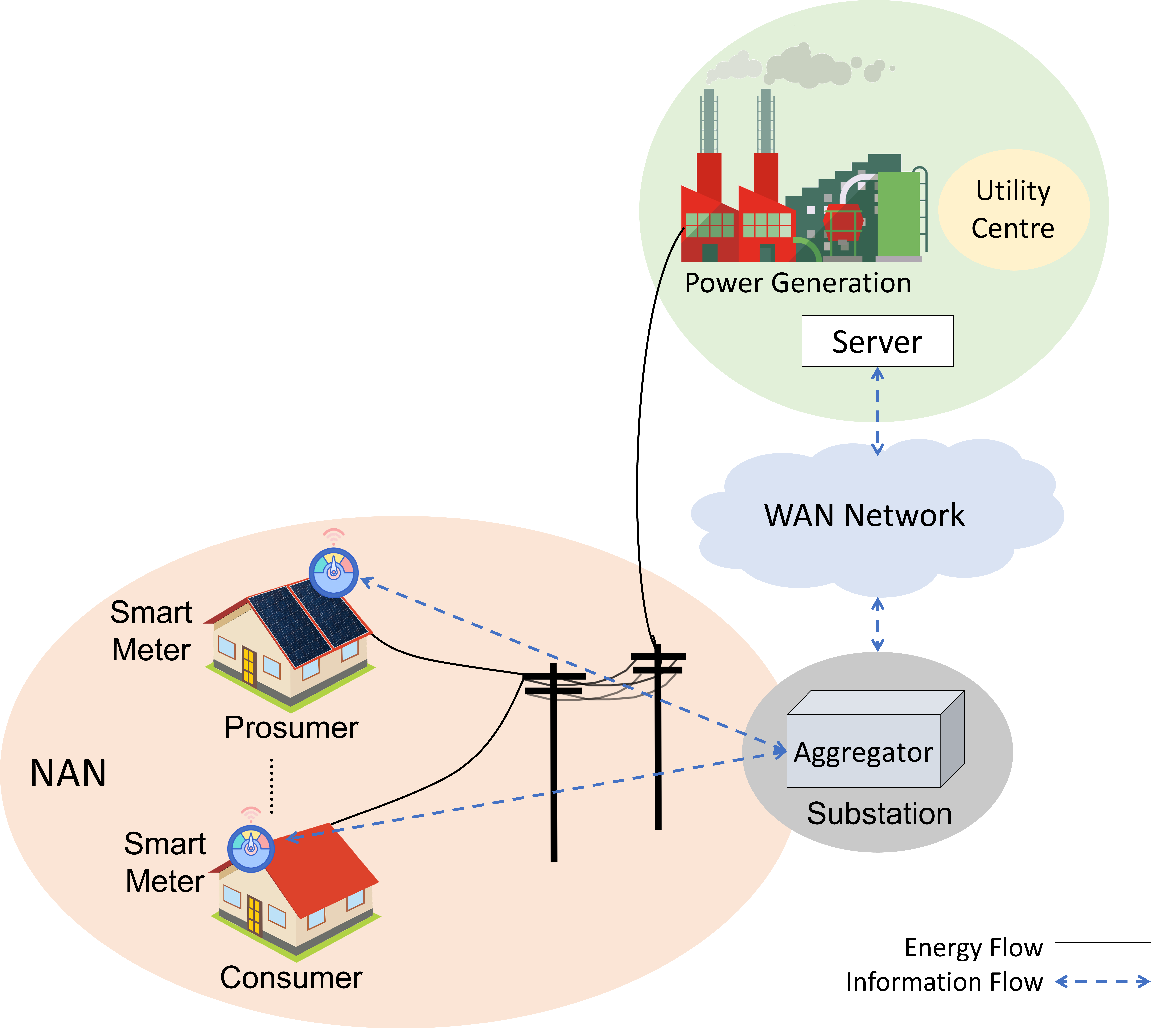}
    \caption{System Model}
    \label{fig:system_model}
    \vspace{-6pt}
\end{figure}

Fig. \ref{fig:system_model} shows our system model that includes three main entities: clients, aggregators, and a server. Specifically, each entity has the following roles in the system:
\begin{itemize}
    \item The server is the utility centre which is responsible for the distribution of electricity to all clients. It is a trusted party that also generates and distributes the masks in our scheme.
    \item Aggregators are third-party components that facilitate the communication and electricity flow between the utility and clients. Each aggregator periodically aggregates the energy measurement data of a group of clients in a geographical location, called a neighbourhood area network (NAN), and sends them to the utility server. The server uses this data to manage the electricity demand for the next period and maintain the balance between power generation and demand.
    \item Clients are homeowners with smart meters that send data to the aggregator in fixed intervals (e.g. every 15 minutes). A client can be either a consumer or a prosumer who both consumes electricity but also produces electricity using a distributed energy resource such as solar panels. Clients are the entities of our proposed system whose data privacy we aim to protect. 
\end{itemize}

\subsection{Threat Model}
In our threat model, we consider a set of threats that includes several possibilities for energy theft attacks and feature inference attacks.
\subsubsection{\textbf{Energy Theft Attacks (ETA)}}\label{ETA} in these attacks, we consider that a percentage of clients may modify their consumption/generation data with the goal of gaining a monetary advantage. This means that an energy theft attack can be launched by either increasing the generation readings to gain extra money or decreasing the consumption readings to lower the bill. Therefore, any deviation from the actual value of both generation and consumption readings is considered energy theft.
This type of attack not only causes monetary losses but also has a huge effect on the safety and accuracy of the whole system's operations. A great number of demand management approaches rely on the data sent by the clients. If there is an energy theft of any kind, it will affect the accuracy and reliability of any estimated future demand. 

Energy theft attacks cover a range of cases that we outline as follows:
\begin{itemize}
    \item \textbf{Case 1 (ETA1):} A consumer may wish to reduce their reported consumption by either a constant value or percentage for a period of time $T$. 
    \item \textbf{Case 2 (ETA2):} A prosumer may wish to increase their reported production by either a constant value or percentage for a period of time $T$. 
    \item \textbf{Case 3 (ETA3):} A consumer may wish to reduce their reported consumption  by a constant value or percentage for a period of time $T$ and increase another consumer's reported consumption by the same energy amount. This is done to maintain the total energy consumed and reported by those two users. This type of attack is 
    called a “balance attack" and was introduced in \cite{alromih2021electricity}.
    \item \textbf{Case 4 (ETA4):}  A prosumer may wish to increase their reported production by a constant value or percentage for a period of time $T$ and decrease another prosumer's reported production by the same energy amount. This is done to maintain the total energy produced and reported by the two users. This is also a 
    balance attack in terms of production. 
    \item \textbf{Case 5 (ETA5):} A prosumer may wish to reduce their reported consumption by a constant value or percentage for a period of time $T$ and increase their own reported  production by the same value. This is a new variant of the balance attack concept launched by a single user.  
\end{itemize}

\subsubsection{\textbf{Feature Inference Attacks (FIA)}}
In our system, we consider the server to be a trusted organisation (e.g. the National Grid in the United Kingdom), whereas most clients and all aggregators are honest-but-curious parties. We call these honest-but-curious entities as \emph{passive adversaries} since they correctly follow the steps of the proposed model but try to passively perform \emph{feature inference attacks} on data outputs that other clients send.  
Feature Inference Attacks (FIAs) have caught the attention of privacy and security researchers after the widespread adoption of collaborative machine learning models in different applications. In an FIA, an attacker causes a target machine-learning model to leak private features (or attributes) from its training data. This means that the adversary tries to find attributes that are close to their true values with a success rate significantly greater than a random guess. 
This has been studied in \cite{yeom2018privacy,zhao2021feasibility,mehnaz2022your,jia2020defending} and by many others in the machine learning community. However, it has not been studied or considered in the area of privacy-preserving machine learning energy theft detection. 

\begin{figure}[tb]
    \centering
    \includegraphics[width=0.9\linewidth]{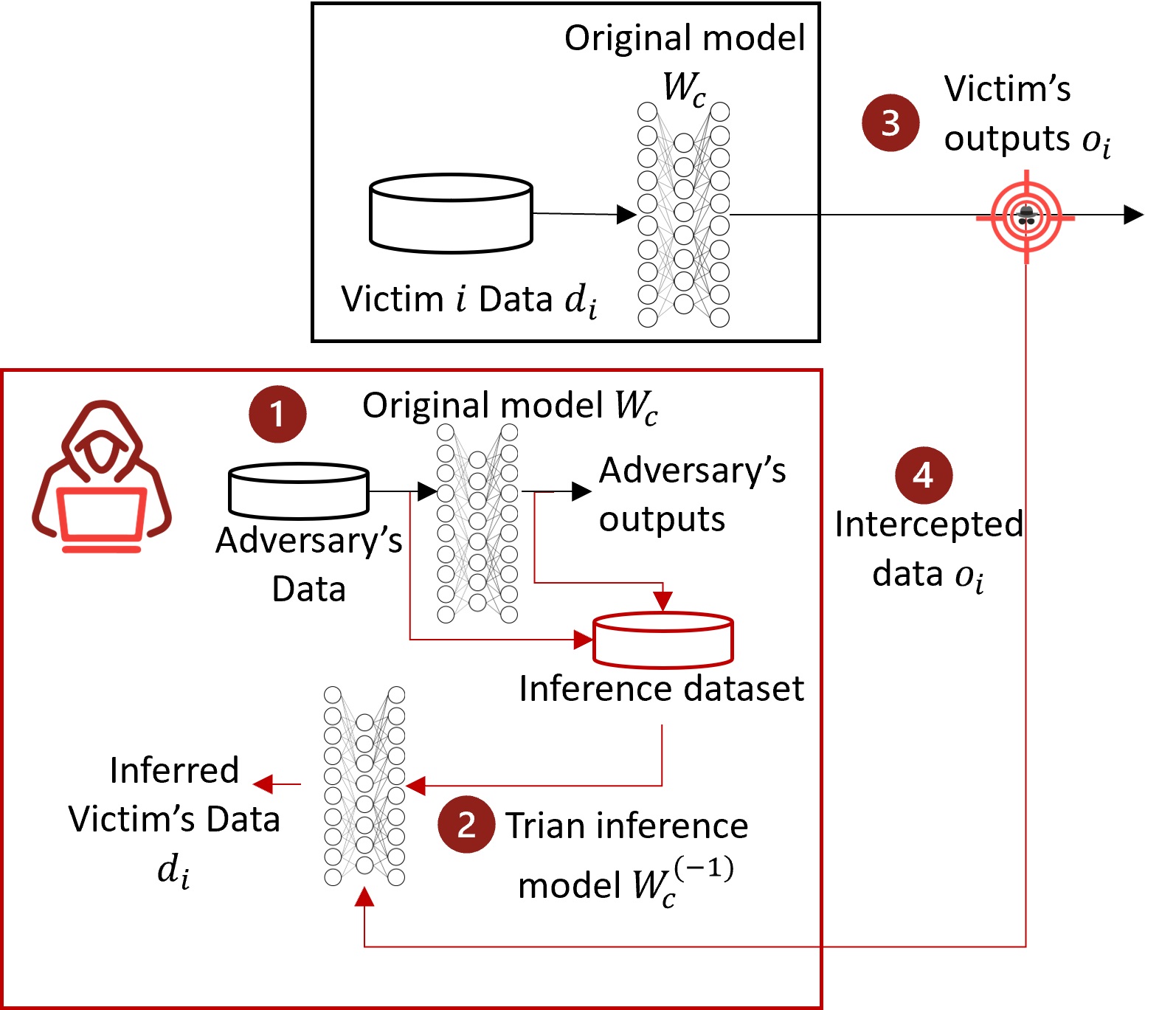}
    \caption{Illustration of the Feature Inference Attack}
    \label{fig:inference_attack}
    \vspace{-6pt}
\end{figure}

In this paper, we view an FIA as an adversarial attack launched by passive adversaries trying to build an inference model that maps the split layer outputs to the original raw readings of the client. 
In particular, the adversary follows these steps to perform the feature inference attack: 1) The adversary builds a dataset that contains their raw readings as targets and the split layer's output as features of this dataset. 2) The adversary builds an inference model $W_c^{(-1)}$ that has the opposite structure of the client's proposed model $W_c$ and trains it to map the client's outputs $o_i$ to their original raw data $d_i$. 3) A victim smart meter $i$ reports their model's outputs $o_i$ (i.e., the output of the split layer via running the forward pass of the victim's model) to the aggregator. 4) The adversary captures those data and tries to infer the original features from the split layer outputs using  $W_c^{-1}$, the previously built inference model from step 2. The steps of this attack are illustrated in Fig. \ref{fig:inference_attack}

We study this attack under three sets of  assumptions:
\begin{itemize}
    \item \textbf{Case 1 (FIA1):} In this adversary setting, we assume that the adversary is a client with a dataset that comes from the same distribution as the victim's training data. Also, since both the victim and the adversary are clients, they have the same model structure. The adversary can use this to their advantage to build the inference model. This is the most strict but realistic attack setting out of the three FIAs.
    \item \textbf{Case 2 (FIA2):} For this setting, we assume that two or more clients collude with each other in an attempt to train a more robust inference model. The inference model is then used to infer features from the split layer outputs of other victims. This adversary setting is more powerful than the previous one as it involves training the inference model with more data, allowing faster convergence. 
    \item \textbf{Case 3 (FIA3):} This is another collision attack that includes the aggregator along with the clients. In this setting, the colluding clients train the inference model using their joint datasets, and the aggregator helps to guess the mask as it already knows the sum of all clients' masks. 
    
\end{itemize}

\noindent Note that the proposed feature inference attack can be performed during the training or deployment phases of the proposed model. 

\subsection{Split Learning and Three-Tier Split Learning}
Split Learning (SL), or SplitNN, is a machine learning technique for training neural networks in a distributed manner without exchanging raw data with the participating entities. SL was first introduced by MIT in 2018 \cite{gupta2018distributed}. In the classical version of split learning, a complete NN model is split into two sections between the client and the server. The input data is processed by the first part of the model on the client side, while the remaining part of the network is on the server side. In SL, only the outputs of the split layer (also called the cut layer) are sent between the entities, unlike in federated learning, where all model weights are shared. 
There are multiple variations to the structure of split learning. One particular variation, called Three-Tier Split Learning, was designed in \cite{alromih2022privacy} to fit the nature of smart grids. In this configuration, an additional entity, i.e. aggregator, is added to the system between the clients and the server. Moreover, split layer activations received from the clients for each client-aggregator pair are averaged by the aggregator before they are used as input for the aggregator's part of the model. Full details about Three-Tier Split Learning can be found in \cite{alromih2022privacy}. We chose this variation for our implementation since it fits our system model.

\subsection{Distance Correlation}\label{distance_correllation}
Distance correlation ($dCor$) is a statistical measure of dependence that was first introduced in \cite{szekely2007measuring}. This measure tests the joint independence between two random vectors $X$ and $Y$ with arbitrary dimensions (lengths). Distance correlation captures linear and nonlinear relationships between the variables, making it a good measure of independence in our case. Specifically, we aim to quantify the dependency between the split layer outputs and the original readings. 
Unlike the classical definition of correlation, in distance correlation, we get a value between 0 and 1 where zero indicates total independence between the two vectors. 
Distance correlation can be calculated by dividing the distance covariance of the two variables by the product of their distance standard deviations. This makes the distance correlation between two random variables $X$ and $Y$ equal to 
\[dCor(X,Y) = dCov^2(X,Y)/ \sqrt{dVar(X)\,dVar(Y)}\]
where $dCov(X,Y)$ is the distance covariance between $X$ and $Y$ and  $dVar(X)$ is the distance covariance between $X$ and itself, i.e. $dCov(X,X)$. The distance covariance $dCov(X,Y)$ is the square root of the average of the product of the double-centred pairwise Euclidean distance matrices and can be calculated as \[dCov^2(X,Y) := \frac{1}{n^2} \sum_{i = 1}^n \sum_{j = 1}^n D(x_i,x_j) D(y_i,y_j)\] 
where the $D(x_i, x_j)$ is the “centred" Euclidean distance between the $i$th and $j$th observations minus the $i$th row mean, the $j$th column mean, and the grand mean of $X$.

\section{Proposed Privacy-Preserving Scheme}
Our proposed privacy-preserving scheme is a multi-output neural network model that takes every client's reading at time $t$ and outputs  three results: an indication of whether theft is suspected or not;  an estimation of the energy theft value, i.e. the deviation from the actual reading (either an increase or a decrease in the production or consumption reading); and the estimated demand of the next period $t+1$. 
In this part, we show how we use split learning along with masking as privacy protection measures to protect the privacy of the client's energy data in the multi-output model. To protect clients' privacy from semi-honest aggregators and eavesdroppers, we employ split learning as the first layer for privacy protection. Moreover, we add an extra privacy-preserving measure, i.e. masking, to split learning to protect the privacy of the client's level outputs and prevent feature-inference attacks. We propose to use a masking-based privacy-preserving scheme that uses pseudorandom number generators to generate masking matrices that mask the clients' outputs of the split layer. The proposed scheme consists of \emph{three} phases: \emph{initialization}, \emph{mask generation and verification}, and \emph{privacy-preserving energy theft detection and demand estimation phase}. A summary of the notations is provided in Table \ref{tab:notations}. For simplicity, we omit the subscript $k$ for every timestamp $t$ as all operations are done in a single timestamp.

\begin{table}[tb]
\small
\begin{threeparttable}
\caption{Notations}
\label{tab:notations}
\setlength\tabcolsep{6pt} 
\begin{tabular*}{\columnwidth}{@{\extracolsep{\fill}} cp{0.8\columnwidth}}
\toprule
\Centering{\textbf{Notation}} & \Centering{\textbf{Description}}\\
\midrule
     $SM_i$ & Smart meter $i$\\
     $Ag_i$ & Aggregator $i$\\
     $M$ & System's modulus\\
     $A_i$ & The multiplier for the pseudorandom number generator algorithm of smart meter $i$ \\
     $C_i$ & The increment for the pseudorandom number generator algorithm of smart meter $i$\\
     $m$ & \# of split layer outputs of a client\\
     $n$ & \# of smart meters in a cluster\\
     $r_i$ & The vector of random numbers used by smart meter $i$\\
     $r_{ji}$ & A random number to mask the $j$th output of smart meter $i$\\
     $d_i$ & The readings (data) of smart meter $i$\\
     $o_i$ & The outputs of the split layer of smart meter $i$\\
     $o_a$ & The output of the aggregator\\ 
     $\hat{o}_i$ & Masked outputs of smart meter $i$\\
\bottomrule
\end{tabular*}

\smallskip
\scriptsize
\end{threeparttable}
\vspace{-6pt}
\end{table}

\subsection{Initialization phase} \noindent This phase consists of the following steps: 

\textbf{Step I1:} During the initialization of the system operations, the server chooses a modulus $M\gg0$ that is used during all system operations. Moreover, each client's smart meter $SM_i$ is initialised with  random parameters $A_i$ and $C_i$ to be used in the pseudorandom number generator algorithm, where the multiplier $A_i$ should be $0 < A_i < M$ and the increment $C_i$ should satisfy $0 \leq C_i < M$. These random parameters are stored on the server side and should be unique for every client. This is because we want to reduce the possibility of generating the same random numbers and recovering the original information in cases where passive adversaries intercept the communication.

\begin{figure}[tb]
    \centering
    \includegraphics[width=0.95\linewidth]{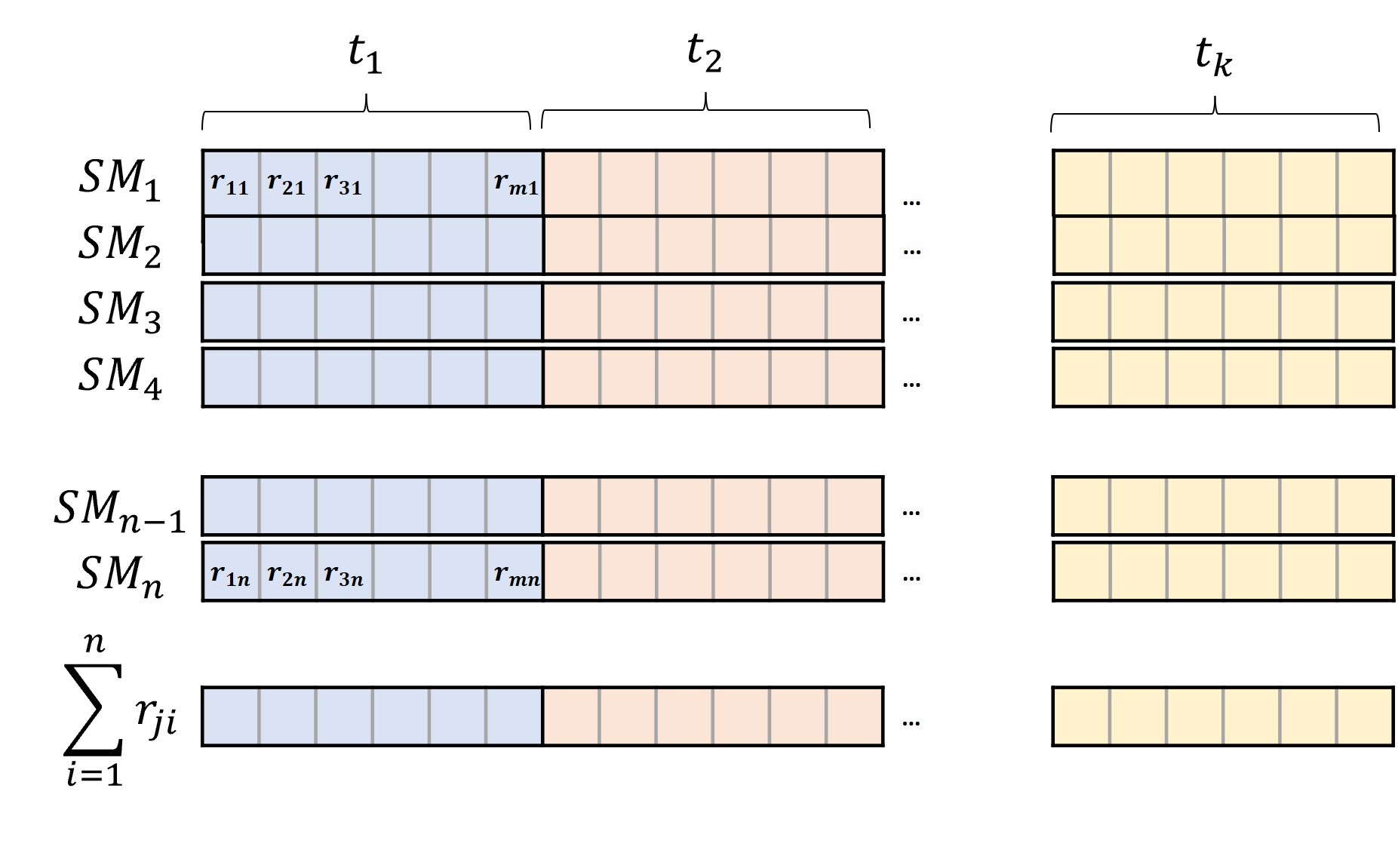}
    \caption{Masking Matrix}
    \label{fig:masking_matrix}
    \vspace{-6pt}
\end{figure}

\textbf{Step I2:} The server initialises each smart meter $SM_i$ with a vector of random numbers of size $1\times(m \times t)$ to be used to mask the $m$ outputs of the split layer for $t$ timestamps. 
This random vector is also stored at the server side for every $SM_i$, which makes the server store a $(n \times (m\times t))$ matrix at its side. Each random number in the matrix is denoted as $r_{ji}$ for each $j \in 1...m$ and $i \in 1...n$. The value of the random numbers should be much larger than the outputs of each neuron to ensure that sensitive information is not leaked.
The full matrix stored on the server side can be seen in Fig. \ref{fig:masking_matrix}, where each row belongs to a single smart meter.

\textbf{Step I3:} The server sends the summation of all random masks $\sum_{i=1}^{n} r_{ji} \ , \ \forall j \in 1...m $ to the aggregator using a secure channel to be used later in the unmasking process.

\begin{figure}[b]
    \centering
    \includegraphics[width=\linewidth]{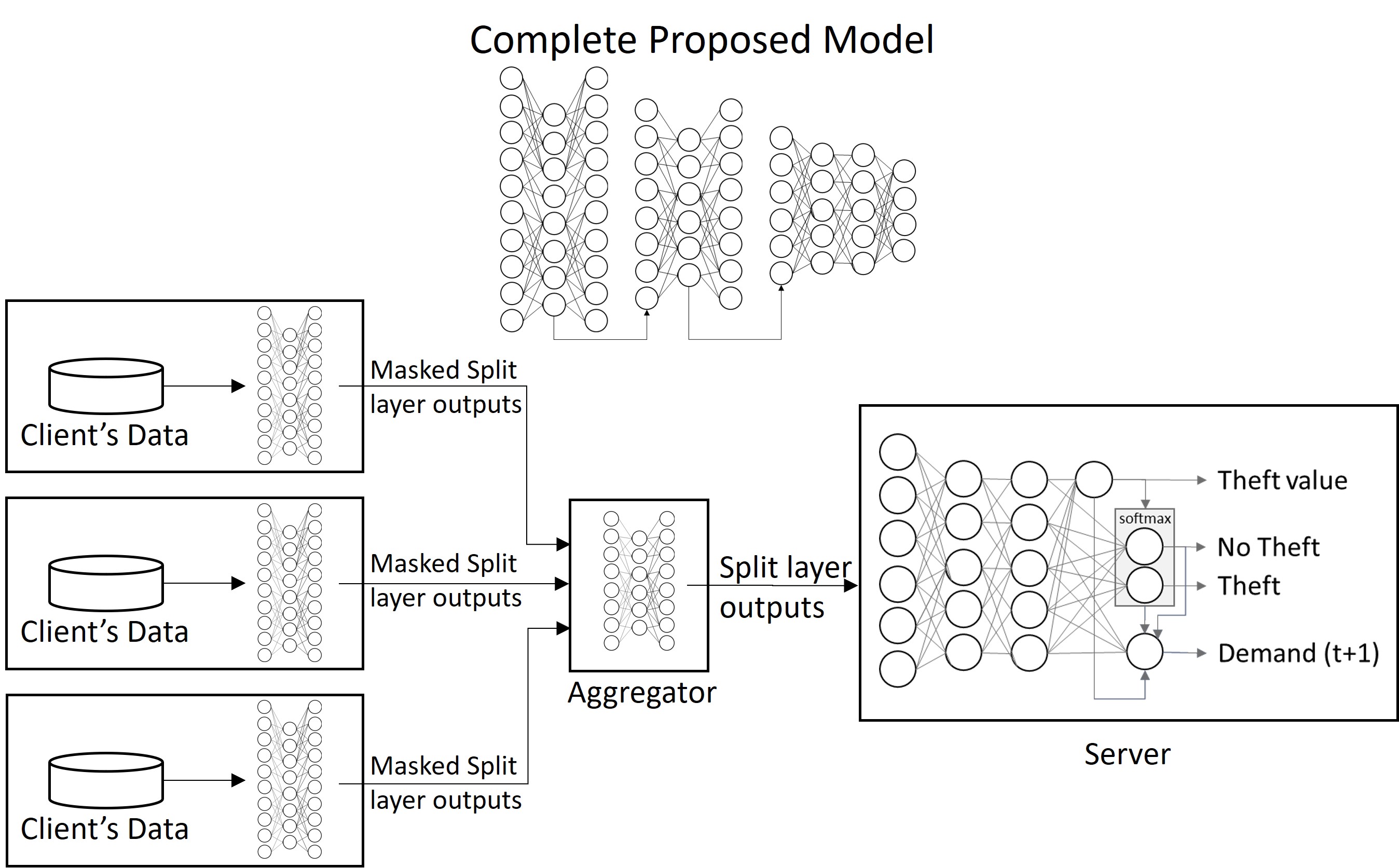}
    \caption{The multi-output NN model architecture}
    \label{fig:splitted_model}
\end{figure}

 \textbf{Step I4:} The multi-output model is split between the system's entities, where every smart meter and every aggregator has a copy of the initialised version of its part. The proposed model is built as a stacked autoencoder (SAE), which is built by stacking multiple autoencoders to extract features layer by layer to obtain deeper and more abstract features that transform sensitive information into non-sensitive abstract data \cite{yin2021stacked,malekzadeh2018replacement}. SAEs have also been proven to be better at producing features than the traditional deep auto-encoders \cite{Wicht2017Deep}. 
The distinct splits of the clients, aggregators and server can be seen in Fig. \ref{fig:splitted_model}. 

\subsection{Mask generation and verification phase}
After $t$ timestamps, the mask vectors and matrix get updated when all the random numbers are used. This is done during the mask generation and verification phase following these steps:

\textbf{Step M1:} Each SM uses its pseudorandom generator parameters $A_i$ and $C_i$ along with the linear congruential generator (LCG) algorithm to generate a new vector of random masks as follows: \(r_{ji}^* = (A_i(r_{ji})+C_i)\ mod \ M)\:, \ \forall j \in 1...m \:\) and \(\forall i \in 1...n\), where $r_{ji}^*$ refers to the new random mask. 

\textbf{Step M2:} At the server side, each client's vector is updated using the LCG algorithm and each client's unique pseudorandom parameters. 

\textbf{Step M3:} The client calculates a hash integrity output $v_i = h(r_{ji}^*\: |\: t \:|\: SM_{id})$ using a one-way hash function and sends $v_i$ to the server to acknowledge that they have the same set of random numbers. 

\textbf{Step M4:} The server validates its sets of random numbers and sends an acknowledgment to the clients to confirm the new set.

\textbf{Step M5:} The summation of each new random mask $\sum_{i=1}^{n} r_{ji}^* \: , \ \forall j \in 1...m $ is sent by the server to the aggregator using a secure channel.


\subsection{Privacy-preserving energy theft detection and demand estimation phase}\label{ETDphase}
This is the main phase of the proposed model, where the proposed multi-output model is trained in a privacy-preserving approach. This phase consists of the following steps: 

\textbf{Step P1:} After the client and the server approve the masking matrix, each client uses their smart meter's energy reading for timestamp $t$ to train their part of the model up to the split layer where they get the outputs $o_i$. 

\textbf{Step P2:} To protect the privacy of these outputs, the client uses the random vector $r_i$ to mask them. However, since the masking is carried out in the integer domain and the client outputs are of floating point numbers, the client needs to quantize the outputs first before masking them. 
Quantization is mainly done to map floats to integers. Specifically, it is done by mapping the min/max of the outputs (weights or activations) with a chosen min/max threshold of the integer range [$-\beta$, $\beta$].
Therefore, the outputs of the split layers $o_i$ are first mapped to an integer in the [$-\beta$, $\beta$] domain by performing the following: $o_i = truncate(o_i \times \beta)$ and then masked with the random vector where $\hat{o}_i = (o_i + r_i) \:mod\: M$ before they are sent to the aggregator.

\textbf{Step P3:} The aggregator aggregates all the masked data received by its set of clients $\sum_{i=1}^{n}\hat{o}_i$. Subsequently, the aggregator needs to unmask the aggregated data by subtracting the summation of the masks. This is done to obtain the average of the unmasked outputs by $Avg(o_i) = (\sum_{i=1}^{n}\hat{o}_i - \sum_{i=1}^{n}{r}_i) \: mod \: M \div n $. The obtained output needs to be dequantized using the opposite quantisation operation to use the average of the clients' outputs in the rest of the model.

\textbf{Step P4:} After that, the aggregator completes its part of the model to get its output $o_a$ and sends these outputs to the server. 

\begin{figure}[tb]
    \centering
    \includegraphics[width=\linewidth]{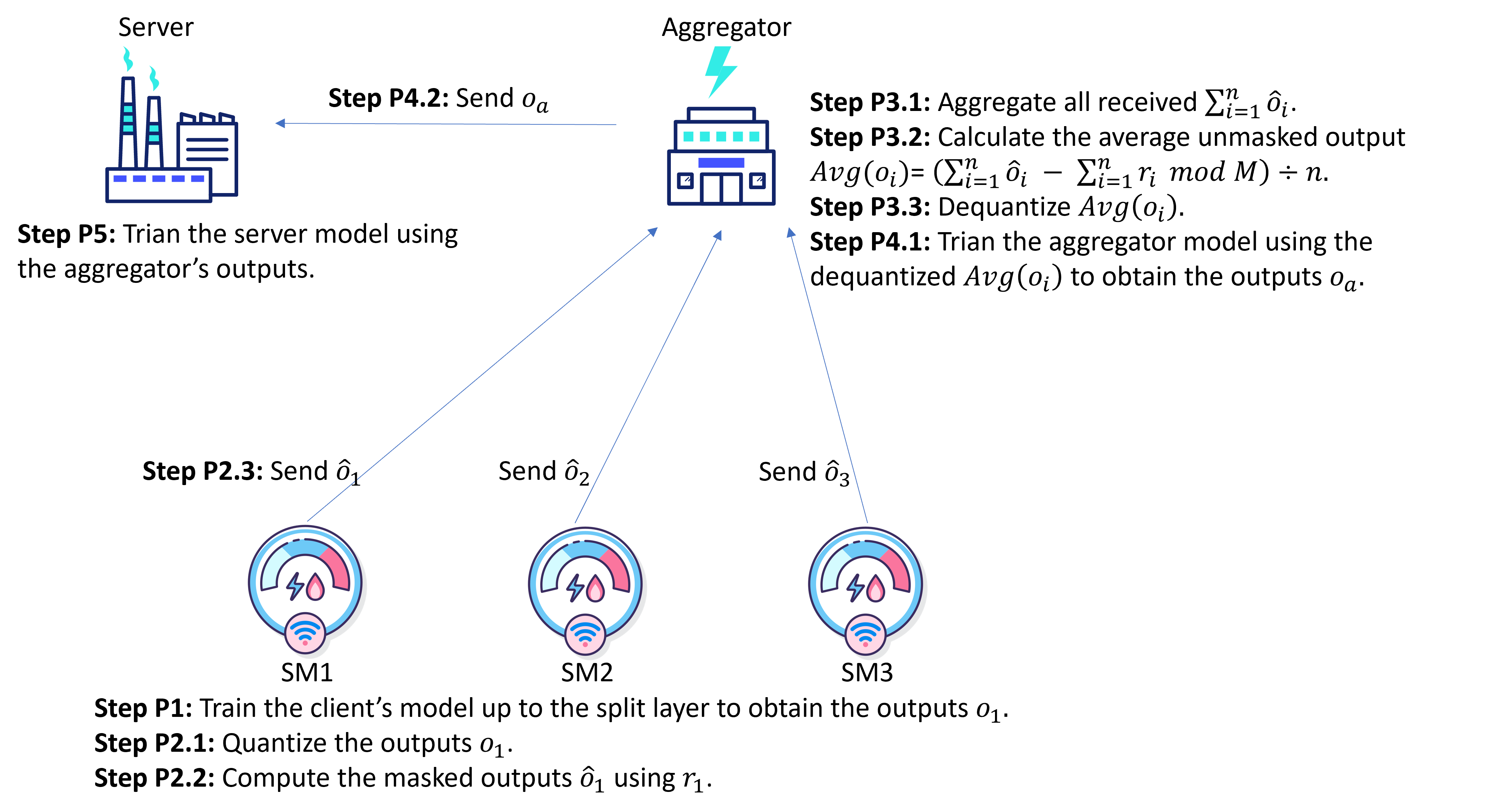}
    \caption{Representation of the privacy-preserving energy theft detection and demand estimation phase steps}
    \label{fig:phase3}
\end{figure}

\textbf{Step P5:} Upon receiving all aggregators outputs at the server side, the server aggregates all received outputs and completes training the ML model. The final output of the server  consists of mainly three outputs: (a) whether each client's input is an energy theft or not. (b) An estimation of the energy theft value. This output estimates how much the reported consumed or produced energy deviates from the actual ones. (c) And the final output estimates the energy demand for the next timestamp $(t+1)$. 
The steps of this phase are illustrated in Fig. \ref{fig:phase3} 

\begin{table*}[!tbh]
\caption{Numerical results of the  proposed scheme for different energy theft attacks}
\begin{center}
\begin{threeparttable}
\setlength\tabcolsep{6pt} 
\renewcommand{\arraystretch}{1.25}
\begin{tabular}{@{\extracolsep{\fill}}
C{0.20\linewidth}rrrrrr}

    \toprule 
     & \centering{\textbf{ETA1}} & \centering{\textbf{ETA2}} & \centering{\textbf{ETA3}} & \centering{\textbf{ETA4}} & \centering{\textbf{ETA5}} & \centering\arraybackslash\textbf{All Theft Types}\\ 
     \midrule
     \textbf{Accuracy} &  99.99 & 99.86 & 99.83 & 99.01 & 99.67 & 94.46\\
     \textbf{Precision}  & 99.99 & 99.85 & 99.79  & 98.97& 99.34 &96.19\\ 
     \textbf{Recall (Detection Rate)} & 99.98 & 99.86 & 99.85 &98.95 & 99.99 & 92.17\\ 
     \textbf{F1 Score} & 99.99 & 99.85 &99.82 & 98.96& 99.66 & 94.14\\       \addlinespace
 
     \hline       \addlinespace

     \textbf{$r^2$ (Theft Value)} & 0.99 & 0.99 & 0.99 & 0.99 & 0.95& 0.92\\ 
     \textbf{$r^2$ (Demand $t+1$)} & 0.86  & 0.84 & 0.82 & 0.85 &0.85 & 0.84\\ 
     \textbf{SMAPE (Theft Value)} & 2.48$\pm$0.04 & 2.89$\pm$0.14 & 3.89$\pm$0.07 & 26.63$\pm$0.64  &  8.39$\pm$0.13  & 8.83$\pm$0.56\\ 
     \textbf{SMAPE (Demand $t+1$)} & 7.57$\pm$0.07 & 7.60$\pm$0.05 & 8.30$\pm$0.07 &  7.87$\pm$0.05  &  7.91$\pm$0.07 & 8.10$\pm$0.06\\ 
      \bottomrule
\end{tabular}
\end{threeparttable}
\end{center} 
\label{tab:ETA_results}
\end{table*}

\subsection*{\textbf{Privacy-enhanced energy theft detection and demand estimation scheme}}\label{privacy_enhanced_model}
We expand our privacy-preserving multi-outputs model (shown in Section \ref{ETDphase}) and propose a more enhanced privacy scheme that trains an additional noisy layer at the client side. The trained client's part of the SAE extracts abstract features from the user's raw data. A small perturbation is added to these abstract features to maximise the independence between these abstract features and the raw data. This is done by training an extra layer at the end of the client's part of the model that takes the client's output as input and outputs the client's output with added Gaussian noise. The loss function of this noisy layer is the distance correlation $dCor$ between the raw data and the noisy output, and the training objective is to minimise it as much as possible. 
The steps of this scheme are as follows:

\textbf{Step PE1:} Define a noisy neural network layer with an input and output size of $m$, the same size as the output of the client's split layer. 

\textbf{Step PE2:} Process the client's raw data through the client's model up to the split layer to obtain $o_i$.

\textbf{Step PE3:} Generate a set of Gaussian noise values and input it to the noisy layer to get a set of noise values added to $o_i$ before they are masked.

\textbf{Step PE4:} Train the noisy layer to add Gaussian noise to the client's output in a way that minimises the $dCor$ between the raw data and the noisy outputs.  This makes the loss function of the whole detection model to be performed in two steps:  the first loss function \textbf{$L1$} is the original loss of the SAE at the server side that is responsible for accurately detecting energy thefts and estimating theft's amount and energy demand, while the second loss function \textbf{$L2$} is minimising the $dCor$ value between the raw data and the noisy outputs. Note that since the two losses have conflicting objectives (accuracy vs privacy), we need to combine the losses into one total loss. The total loss function can be calculated as follows: $Loss Function = L_1 + \alpha L_2$, where $\alpha$ is the trade-off percentage between accuracy and privacy. 
The remaining procedure in this scheme is carried out similarly to how they are in Section \ref{ETDphase}, starting with Step P2.

\section{Experiments and Results}
We conducted several experiments to cover the two broad threats identified in our threat model. In the first set of experiments, we evaluate the accuracy of our proposed model in detecting energy thefts and estimating theft values and energy demand. This set of experiments highlights results in cases of attacks \emph{ETA1} to \emph{ETA5}, explained in our threat model. In the second part of this section, we evaluate the privacy of our proposed model and the privacy-enhanced version in terms of distance correlation and the successfulness of the three feature inference attacks (\emph{FIA1} to \emph{FIA3}) explained in our threat model. Results of the above two sets of experiments are presented in Section \ref{ETA_experiments} and Section \ref{privacy_experiments}, respectively. Finally, in Section \ref{comparison_expermint}, we compare the proposed privacy-preserving approach with state-of-the-art approaches to give a fair view of where our proposed scheme stands.

\subsection{Experimental Setup}
To evaluate our proposed scheme, we built our multi-output neural network using multiple Python 3 \cite{python_2009} libraries which are Pandas \cite{mckinney2010data}, PyTorch \cite{pytorch_2019} and Optuna \cite{optuna_2019}. We used Pandas for preprocessing the data, PyTorch for implementing the three-tier split learning architecture, and Optuna was used as a hyperparameter optimization framework to automate the search for the optimal hyperparameters for our proposed scheme. The search space included four different hyperparameters, including the number of hidden layers (between 6 and 20 inclusive), each hidden layer size (between 4 nodes and 128 nodes inclusive), the learning rate (between $1e^{-1}$ to $1e^{-5}$ inclusive), and the optimizer algorithm (either Adam, SGD or RMSprop). This search space was evaluated with a multi-objective function of maximising the accuracy of energy theft detection and minimising the mean squared error of both the estimated energy theft and the estimated demand. Preliminary trials using Optuna furnished us with the following effective hyperparameters used throughout our experiments: 10 hidden layers of sizes $[65,91,33,89,72,33,76,30,56,44]$, a learning rate of $1e^{-4}$ and with the Adam optimizer. The ten hidden layers are split between the three tiers: client, aggregator, and server, as three layers, three layers, and four layers, respectively.

For our experiments we used our own developed dataset \cite{alromih2021electricity}. This includes energy readings of 1596 clients, 49 of which are prosumers with solar panels. For each client, 14 physical features are reported on each property. These features are  floor area size, number of stories, ceiling height, thermal integrity levels for the floors, walls, doors, and roofs, number of glazing layers, glazing treatment, glass type, windows frame type, heating and cooling system types, and solar panel size. In the dataset, readings of 16 dynamic features are reported every 15 minutes,  which include electricity parameters (power consumption, power generation, voltage, current, reactive power, and apparent power) and weather parameters (temperature, wind speed, wind direction, pressure, humidity, solar radiation, extraterrestrial radiation, direct and diffuse horizontal radiation, solar illumination, and sky cover). This dataset has been made available on our GitHub repository\footnote{https://github.com/asr-vip/Electricity-Theft}.

Since the readings in this dataset are all true (normal) readings, we mathematically changed 50\% of them to be malicious data points. This was first presented in \cite{jokar2015electricity} and is frequently used in most energy theft detection research. The attack scenarios described in Section \ref{ETA} are used to produce these malicious points where constant deviations are randomly chosen between 100 and 400 watts, and percentage deviations are also randomly chosen between 10\% and 40\% of the actual reading.
To process the dataset, we split it into 80\% training and 20\% testing with a batch size of 128. We also normalised all input features using the Min-Max scaler using the default range [0,1] and extracted each reading's minute, hour, month, day, day of the year, and day of the week as extra features from the timestamp.

\subsection{Energy Theft Detection Experiments}\label{ETA_experiments}
The results of our energy theft attack detection and energy theft value and demand estimations are promising. We evaluated the performance of our system using several metrics, including accuracy, precision, recall (detection rate DR), and F1 score. We also used the coefficient of determination $r^2$ and the symmetric mean absolute error $SMAPE$ with 95\% confidence level to evaluate how good our estimations are for both estimating the theft value and the demand for the next timestamp $(t+1)$. These results are shown in Table \ref{tab:ETA_results}, where we reported the results of every energy theft attack type presented in Section \ref{ETA} on a single column. Then all attack types are presented together in a single dataset in the last column, \emph{“All Theft Types"} column. The table also shows the $r^2$ and $SMAPE$ results for our two outputs estimations.

\begin{figure}[tb]
     \centering
    \subfloat[Fraudulent Deviation of Consumed/Produced Energy\label{fig:theft_value}]{%
         \centering
         \includegraphics[width=0.41\linewidth]{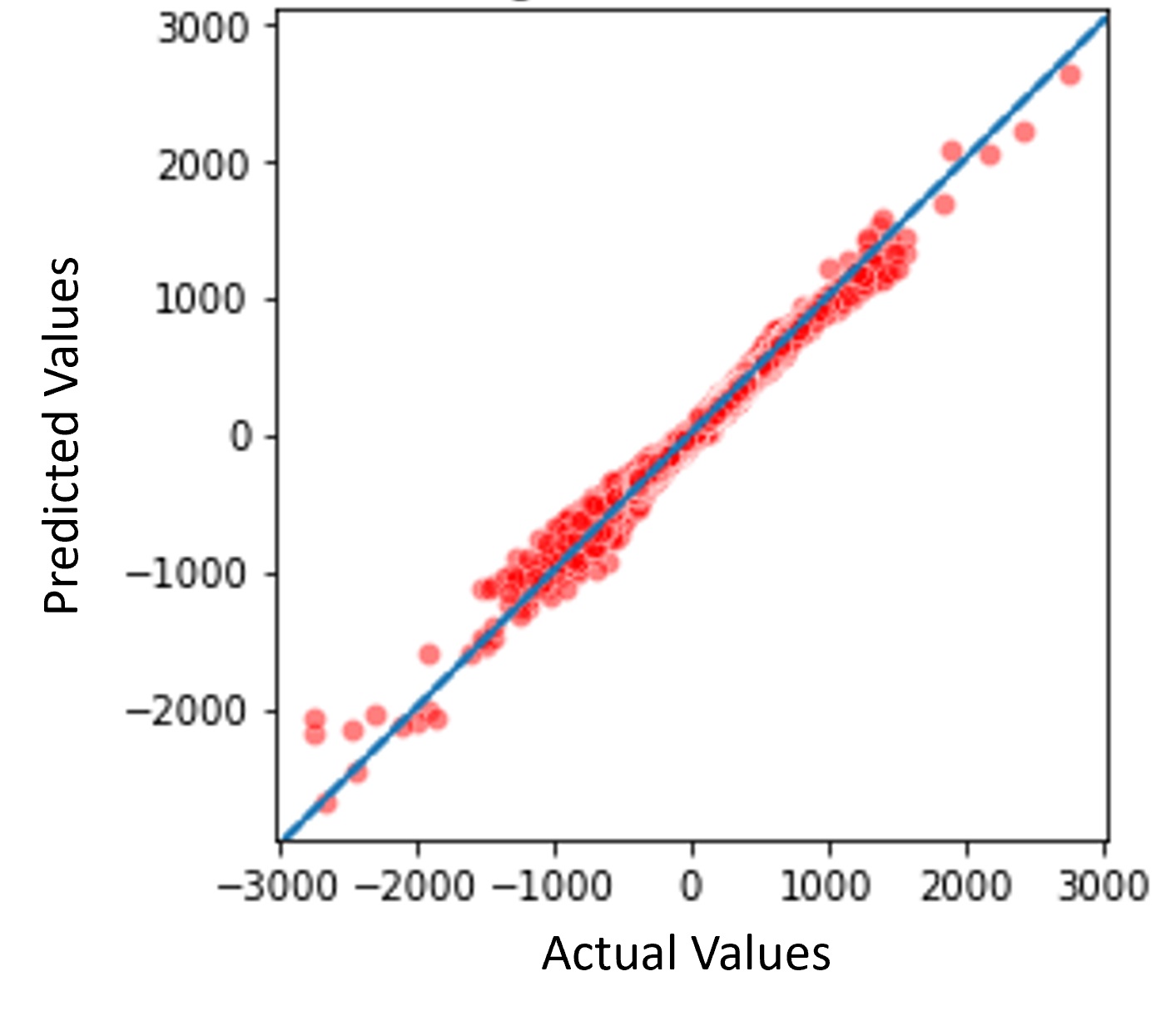}}
     \hfill
          \subfloat[Demand ($t+1$)\label{fig:demand}]{%
         \centering
         \includegraphics[width=0.57\linewidth]{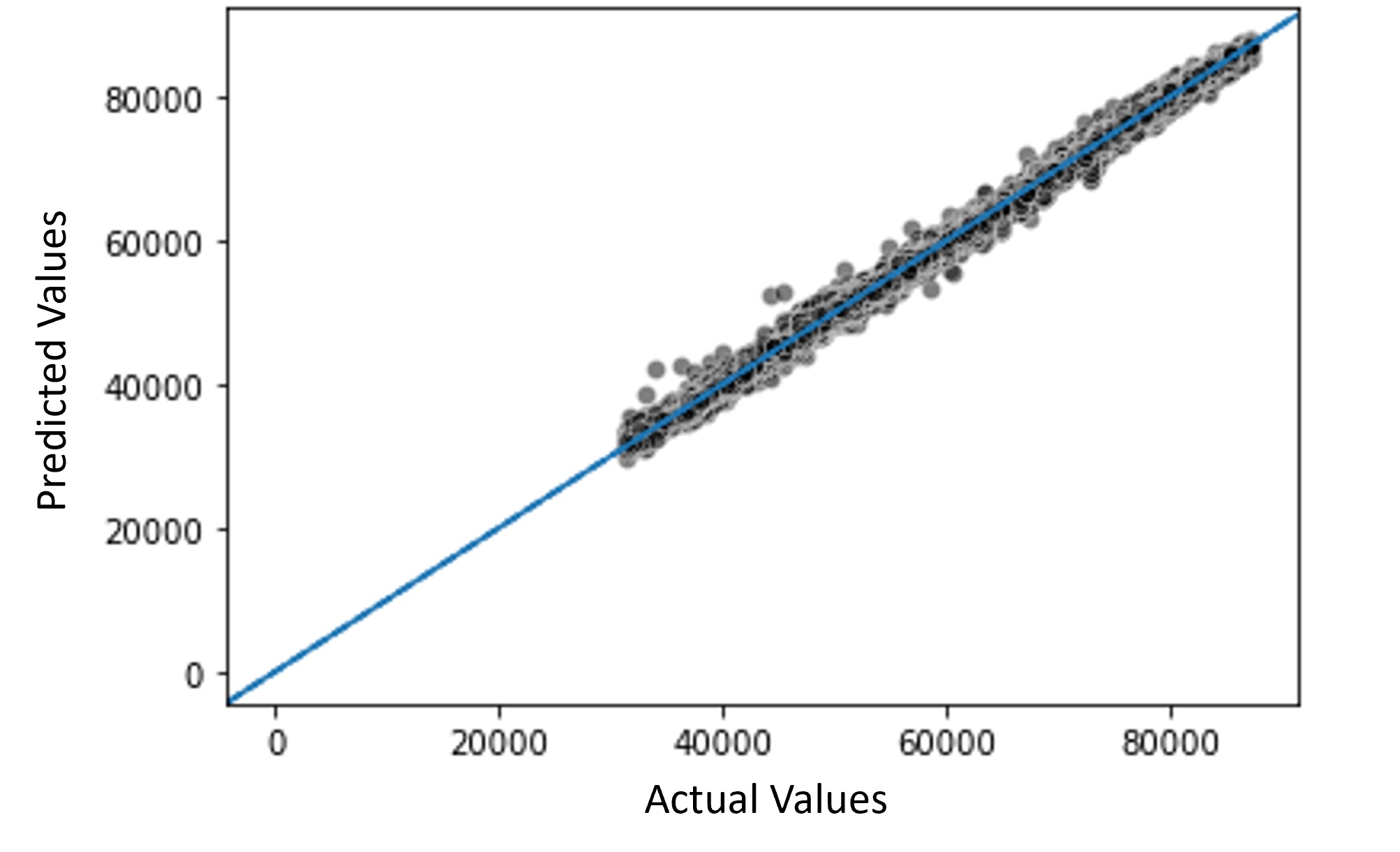}}
        \caption{Actual vs. Predicted values of the model's two outputs: Demand ($t+1$) and Theft Value (Fraudulent deviation of consumed/produced energy }
        \label{fig:demand and theft_value}
        \vspace{-6pt}
\end{figure}

From the results in Table \ref{tab:ETA_results}, we can see that our detection model performs exceptionally well in detecting all types of energy theft attacks. The overall performance is very good in the case of “All Theft Types" with an accuracy and F1 score of about 94\% for both, a precision of 96.10\%, and a detection rate of 92.17\%. Moreover, our model performs well in estimating the energy theft values and the demand for $(t+1)$, which is also illustrated in Fig. \ref{fig:demand and theft_value}. The table shows that the $r^2$ of the estimated theft value and the demand $t+1$ ranged between 0.99 and 0.82, indicating good performance. The SMAPE values from the table show the percentages of the mean absolute error for our two estimates. We can see that the average percentage of error in estimating the theft value of “All Theft Types" is equal to 8.83\%$\pm$0.56 with a 95\% confidence level. In particular, in the event when all theft types are present in the dataset, our theft value estimates are between  (the actual value - 8.83\%) and (the actual value + 8.83\%), i.e. 
\(\textrm{Actual value} - 8.83\% \leq \textrm{Predicted value} \leq \textrm{Actual value} + 8.83\%\). 
Furthermore, the estimates of the future demand for ($t+1$) are better with an absolute percentage error of 8.10\% for the “All Theft Types" case. 
These results are also confirmed in Fig. \ref{fig:theft_value} and Fig. \ref{fig:demand}, where the actual values of theft (fraudulent deviation of consumed/produced energy) and demand are plotted against the predicted ones.  From the two sub-figures, we can see that the model is performing well. The predicted values are significantly close to the regression line (i.e., the actual values), indicating low percentages of error.

\begin{figure}[b]
     \centering
    \subfloat[Performance of estimating  demand ($t+1$) taking theft detection and theft value estimation into account\label{fig:demand_experment1}]{%
         \centering
        \includegraphics[width=0.49\linewidth]{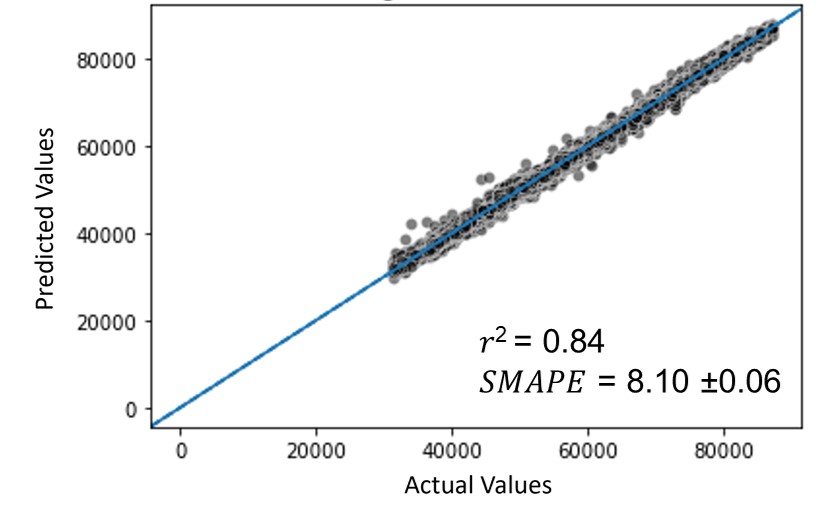}}
     \hfill
          \subfloat[Performance of estimating demand ($t+1$) without taking theft detection and theft value estimation into account\label{fig:demand_experment2}]{%
         \centering
         \includegraphics[width=0.49\linewidth]{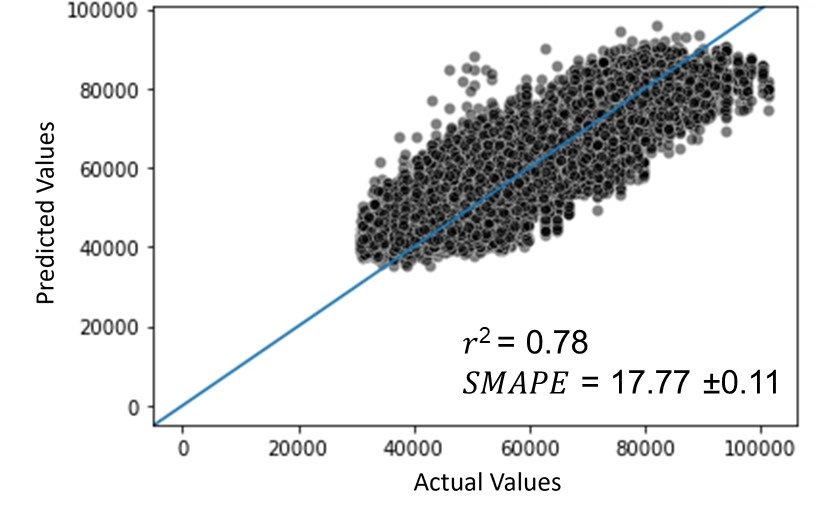}}
        \caption{Performance of demand-response management part of the system in two cases: (a) taking theft detection and theft value estimation into consideration, and (b) without taking them into consideration}
        \label{fig:demand-response management}
\end{figure}

To support our main motivation, which states that managing the demand-response of energy needs to take energy theft detection and estimation into account, we performed an experiment where we compared the performance of our demand estimation output in two  cases: (a) taking theft detection and theft value estimation outputs into account and (b) in the case where the previous two outputs do not contribute to the demand ($t+1$) estimation output. The results of the two cases can be seen in Fig. \ref{fig:demand-response management}, where it is observable that considering theft detection and theft value estimation when estimating the future demand has a great advantage. There is an improvement of around 7.7\% in how well our estimator  estimates the actual demand values (i.e. $r^2$ value) and a significant decrease in the symmetric mean absolute percentage error ($SMAPE$) of the demand's estimates from 17.77 $\pm$0.11 to 8.10 $\pm$0.06. This confirms that taking theft detection and the estimated theft values into consideration in demand-response management is beneficial and justified.

\begin{remark}
It is important to consider thefts and their values in managing the future demand for an energy system.
\end{remark}

\begin{table*}[!ht]
\caption{Performance of different masking levels $\beta$ and different noisy layer training levels $\alpha$}
\renewcommand{\arraystretch}{1.25}
\begin{center}
\begin{threeparttable}
\setlength\tabcolsep{0pt} 
\begin{tabular}{@{\extracolsep{\fill}}
@{\hspace{6pt}}p{0.21\textwidth}p{0.08\textwidth}@{\hskip 12pt}p{0.09\textwidth}p{0.09\textwidth}p{0.09\textwidth}p{0.09\textwidth}p{0.16\textwidth}}
    \toprule
\multicolumn{2}{c}{\textbf{Scheme Used}}& {\textbf{Accuracy}} & {\textbf{Precision}} & {\textbf{Recall (DR)}} & {\textbf{F1-Score}} & {\textbf{Distance Correlation}}\\ \midrule
     \multicolumn{2}{l}{\hskip6pt Non-privacy-preserving approach (no masking)} & 93.00 & 92.75&92.73&92.74& 0.802  \\
     \multirow{8}{0.20\textwidth}{Privacy-preserving proposed scheme (with masking)}& $\beta=10e8$ & 94.46 & 96.19 & 92.17 & 94.14 & 0.612 (-23\%)
     \\ 
       & $\beta=10e7$ & 93.44&	93.56&	92.79&	93.17 & 0.613 \\ 
      &$\beta=10e6$ & 93.05&	93.19& 92.35& 	92.77& 0.613\\ 
      &$\beta=10e5$ & 92.28&	90.48&	93.88&	92.15 & 0.613\\ 
     &$\beta=10e4$ & 91.86&	91.11&	92.10&	91.60& 0.667  \\ 
      &$\beta=10e3$ & 89.36&	86.81& 91.92&	89.29& 0.692\\ 
     &$\beta=10e2$ & 57.53&	53.36&	94.91&	68.31& 0.797\\ 
      &$\beta=10e1$ & 55.39&	52.08&	94.25&	67.08& 0.801\\
      \addlinespace
      \hline
      \addlinespace
    \multirow{9}{0.20\textwidth}{Privacy-enhanced proposed scheme (with masking \& noisy layer)} & \raggedright $\beta=10e8$ $\alpha= 0.01$ & 77.45 & 75.33 & 80.24 & 77.42 & 0.238 (-70\%)\\ 
     & \raggedright$\beta=10e8$ $\alpha = 0.005$ & 86.23 & 90.13 & 82.45 & 85.21 & 0.241 (-70\%)\\ 
     & \raggedright$\beta = 10e8$ $\alpha = 0.001$ & 88.64 & 93.75 & 82.11 & 88.30 & 0.295 (-63\%)\\ 
     & \raggedright$\beta = 10e8$ $\alpha = 0.0005$ & 89.66 & 95.13 & 82.43 & 88.29 & 0.289 (-64\%)\\ 
     & \raggedright$\beta = 10e8$ $\alpha = 0.0001$ & 91.14 & 95.41 & 90.33 & 89.23 & 0.525 (-35\%)\\ 

      \bottomrule
\end{tabular}
\end{threeparttable}
\end{center} 
\label{table:distanceCorrelation1}
\end{table*}

\subsection{Privacy Experiments}\label{privacy_experiments}
To assess how much our proposed model enhances the privacy aspect of the detection approach, we used two different sets of evaluation metrics. The first is by using distance correlation, defined in Section \ref{distance_correllation}, and the other is by measuring the inference error of an inference attack. The inference error shows the degree of accuracy in inferring the private raw features, where higher errors indicate a lower likelihood of successfully launching a feature inference attack. These two metrics are assessed in the following two subsections.

\subsubsection{Distance Correlation}
We evaluate the distance correlation $dCor$ between the users' inputs and the outputs that they send to the aggregator in an attempt to measure both linear and non-linear dependencies between the two. The aim is to lower this dependence as much as possible so that it would be difficult for an attacker to launch a successful feature inference attack.
The first part of Table \ref{table:distanceCorrelation1} compares the results between the non-privacy-preserving approach and the proposed privacy-preserving one using different masking levels. Each row indicates a different case where we used different values for $\beta$, which is the quantization limit.
As can be seen from the table, the $dCor$ value in the non-privacy approach is equal to 0.801, which is high, indicating  a strong dependency between the private SM's inputs and the sent outputs. Then when we apply the privacy-preserving approach, the $dCor$ value decreases to 0.612, improving the privacy levels of the client's private reading by about 23\%. Also, when setting $\beta$ to the maximum level, we get almost the same detection performance results as the non-masking case, with a huge reduction (around 23\% decrease) of the $dCor$ value between the user's input data and the outputs sent to the aggregator. The table also shows that the detection performance is unreliable when we set $\beta$ to values less than $10e3$. The reason is that quantizing a float to an integer with small ranges results in a huge precision loss which leads the model to be unable to learn how to detect theft accurately. Therefore, we adopt the value of $\beta = 10e8$ for all future experiments where we apply our proposed privacy-preserving approach.

The second part of Table \ref{table:distanceCorrelation1}, shows distance correlation results when the privacy-enhanced version of our proposed scheme is used. As described in Section \ref{privacy_enhanced_model}, we add a noisy layer to the proposed approach to help conceal the users' inputs. The table shows how adding this noisy layer with different $\alpha$ values improves the distance correlation results. However, we also see that by setting $\alpha$ too large, the detection performance decreases as the machine learning model tries to optimise the distance correlation more. Setting $\alpha$ to a small value of 0.0001 will still give the same detection performance with increased privacy preservation of around 35\% compared to the non-privacy approach. 

\begin{remark}
    There is a clear trade-off between privacy and detection performance. The better privacy degree we achieve from lowering the $dCor$, the worse the results are in terms of detection accuracy.
\end{remark}

\begin{figure}[tb]
    \centering
    \includegraphics[width=\linewidth]{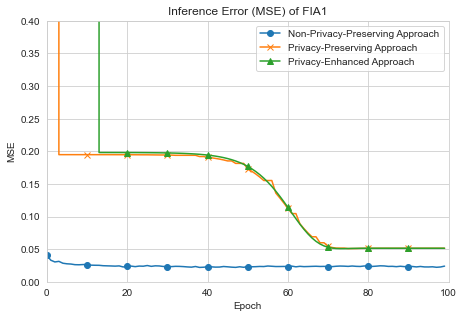}
    \caption{Inference error of inference attack \emph{FIA1} using the non-privacy-preserving approach, the proposed privacy-preserving approach and the proposed privacy-enhanced approach}
    \label{fig:inference_attack_one_client}
\end{figure}

\begin{figure*}[tbh]
    \centering
    
    \subfloat[Non-privacy-preserving approach\label{fig:FIA2_error_non_masking}]{%
        \centering
\includegraphics[width=8.8cm]{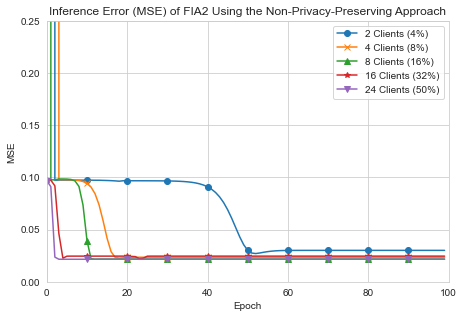}}\hfill
    \subfloat[Privacy-preserving approach\label{fig:FIA2_error_masking}]{%
        \centering
\includegraphics[width=8.8cm]{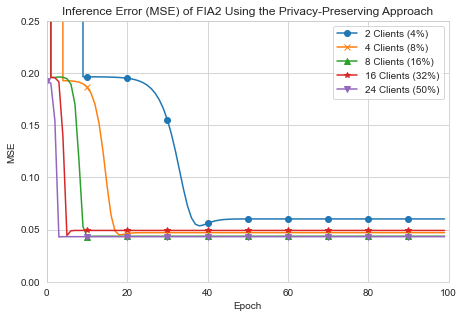}}
\\
    \subfloat[Privacy-enhanced approach\label{fig:FIA2_error_masking_and_noisy_layer}]{%
        \centering
 \includegraphics[width=8.8cm]{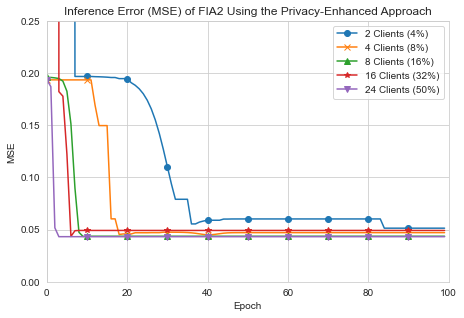}}\hfill       
    \subfloat[Summary comparison between the three approaches\label{fig:FIA2_summarry}]{%
        \centering
\includegraphics[width=8.7cm]{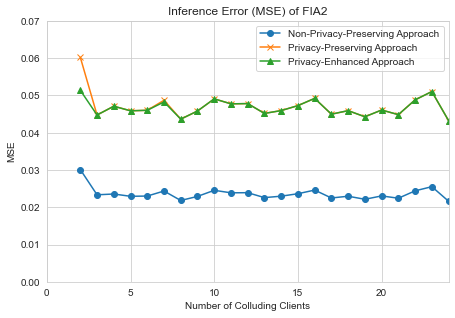}}
        
        \caption{Inference error of \emph{FIA2} using the non-privacy-preserving approach, the proposed privacy-preserving approach, and the privacy-enhanced one.}  \label{fig:FIA2_error}    
\end{figure*}

\subsubsection{Inference Error}
In this set of experiments, we evaluate the attacker(s)' abilities to launch a successful feature inference attack (FIA) against our proposed schemes. We compare how good the attacker is in inferring the victims' original data after building an inference model where these  experiments are done as follows: the attacker(s) build an inference model $W_c^{-1}$ by training an inverted version of the client's original model $W_c$. This inference model $W_c^{-1}$ takes the split layer outputs $o_i$ or the masked output $\hat{o}_i$ as inputs and trains the model to map them to their original data $d_i$. The inference model is trained using the attacker(s)' own data, which means that the more clients that collude to train the inference model, the more powerful it is. The model is tested to infer other victims' data from the outputs they send where these victims' data are not part of the training phase. After that, the mean squared error (MSE) between what has been inferred and the actual data is measured to assess the feature inference attack accuracy rate, where lower values of MSE indicate higher chances of attack success.

\paragraph{FIA1 Experiment}
In our first experiment, we look at the first type of feature inference attacks, \emph{FIA1}, where one attacker builds an inference model and tries to infer other victims' original data. In Fig. \ref{fig:inference_attack_one_client}, we measured the average MSE of the inference model built in three cases: when the non-privacy-preserving approach (no masking) is used, when the proposed privacy-preserving approach (with masking) is used, and finally when the proposed privacy-enhanced approach (with masking and noisy layer) is used.
As we can see from the figure, the inference error is much less when the non-privacy-preserving approach is applied. The error is an average of $\sqrt{0.022} = 0.14$, which is significantly lower than the error in cases where one of our proposed privacy-preserving approaches is applied. With the proposed privacy-preserving and the privacy-enhanced approaches, the error is big in the first training rounds with an average value of $\sqrt{0.2} = 0.44$. This means that the inferred values have a mean error of 0.44, which is very high considering that all of our original raw feature values are normalised in a range between [0-1]. At later stages of the training, after 50 epochs, the error drops to around $\sqrt{0.051} = 0.22$, which is still double the error of the non-privacy-preserving approach.
\begin{remark}
    Using one of the proposed privacy-preserving approaches doubles the error of the feature inference attack making the attack less successful. 
\end{remark}

\paragraph{FIA2 Experiment}
In the second experiment, we assess the inference error in case two or more clients collude with each other, which we refer to as \emph{(FIA2)}. In this attack, the colluding attackers will train an inference model using their joint datasets.  The built model is used to infer other victims' original data (those clients that did not participate in the training).  Fig. \ref{fig:FIA2_error} shows the results of the inference error in terms of MSE in three cases: (a) using the non-privacy-preserving approach; (b) using the proposed privacy-preserving approach; (c) using the proposed privacy-enhanced approach; and finally (d) a summary figure comparing the three approaches together. In all cases, there were 49 clients in the same cluster, and we tested with different percentages of colluding clients. The first thing we notice from these figures is that as the number of colluding attackers increases, the inference error drops faster. This confirms the basics of any machine learning where having more data in training results in faster convergence and more accurate models \cite{peteiro2013survey}.

Looking at Fig. \ref{fig:FIA2_error_non_masking}, we can see the results of performing an \emph{FIA2} attack in the case where the non-privacy-preserving approach is used. Once again, we can see that the inference model converges faster and gives better results as the number of colluding clients increases. Moreover, in all the cases of different colluding clients numbers, the MSE of the inference attack is at around 0.022, which is almost half the error in cases where either the proposed privacy-preserving approach (shown in Fig. \ref{fig:FIA2_error_masking}) is used or when the privacy-enhanced version  (shown in Fig. \ref{fig:FIA2_error_masking_and_noisy_layer}) is used. 
To summarise these results, the last sub-figure, Fig. \ref{fig:FIA2_summarry}, shows a comparison between the inference error of attack \emph{FIA2} using the non-privacy-preserving approach and the two proposed privacy-preserving approaches. In this figure, the MSE is used for this comparison after training the model for 100 epochs. We can see that the inference error of the proposed privacy-preserving approaches is more than the non-privacy approach for all different numbers of colluding clients.  It is double the error in all these cases, indicating an added advantage of using the proposed schemes over the non-privacy one and making it difficult for attackers to launch accurate feature inference attacks even when they collide. 
\begin{remark}
    An increased number of colluding attackers in an FIA allows those attackers to  get a more accurate inference model faster.
\end{remark}


\begin{figure}[tb]
    \centering
    \includegraphics[width=\linewidth]{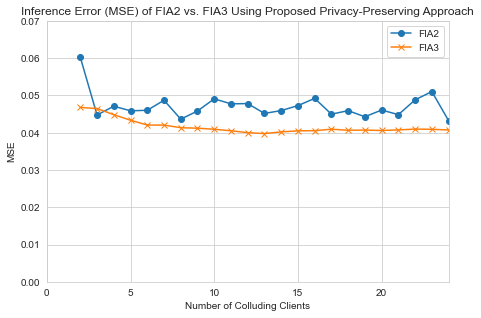}
   \caption{Inference error of FIA3 vs. FIA2 using the proposed privacy-preserving approach}
   \label{fig:FIA3_error}
\end{figure}

\paragraph{FIA3 Experiment} 
In this experiment, we evaluate the attackers' ability to perform the last feature inference attack, \emph{FIA3}. In this attack, a group of malicious clients collude with an aggregator. We performed this attack in case our proposed privacy-preserving approach is used and compared it with the results of FIA2. From Fig. \ref{fig:FIA3_error}, we see that there is not much advantage of having the aggregator as a collaborator in this attack. This proves that splitting our proposed privacy-preserving scheme between the system's entities improves the system's privacy even with honest-but-curious aggregators.

\subsection{Comparison With Other Privacy-Preserving Schemes}\label{comparison_expermint}
To give a fair evaluation of our proposed model against state-of-the-art energy theft detectors, we compared our work with the previously reviewed privacy-preserving, federated-learning-based approaches: FedDetect \cite{wen2021feddetect} and FedDP \cite{ashraf2022feddp}. In particular, Table \ref{tab:ETA_comparision_results} shows the results of the three models in terms of accuracy, precision, recall, and F1-score. As we can see from this table, the performance of the three approaches is relatively the same but with a great increase in the recall (detection rate) result from our proposed scheme. Therefore, we can argue that our approach gives better results in terms of energy theft detection. Moreover, although the performance of our proposed scheme is almost the same as \cite{wen2021feddetect} and  \cite{ashraf2022feddp} in terms of energy theft detection, these two schemes lack some important features, as discussed previously in Section \ref{related_work} and Table \ref{tab:related_work}. Both \cite{wen2021feddetect} and  \cite{ashraf2022feddp} do not provide functionalities for estimating the energy theft value, estimating the future demand, and preserving privacy with minimal communication and computation overheads.

\begin{table}[tb]
\caption{Comparison of previous literature with the proposed scheme in terms of accuracy, precision, recall, and F1 score}
\begin{center}
\begin{threeparttable}
\setlength\tabcolsep{1pt} 
\renewcommand{\arraystretch}{1.25}
\begin{tabular}{@{\extracolsep{\fill}}
C{0.25\linewidth}p{0.15\linewidth}p{0.15\linewidth}p{0.15\linewidth}p{0.15\linewidth}}
    \toprule
     & \centering{\textbf{Accuracy}} & \centering{\textbf{Precision}} & \centering{\textbf{Recall}} & \centering\arraybackslash{\textbf{F1 Score}}\\ 
     \midrule
     \textbf{FedDetect \cite{wen2021feddetect}} &  91.90 & - & - & -\\
     \textbf{FedDP \cite{ashraf2022feddp}}   & 91.67 & 89.03 & 91.67 & 88.72   \\ 
     \textbf{Our Proposed Scheme} & 91.63 & 88.18 & 96.14 & 91.99 \\ 
      \bottomrule
\end{tabular}
\end{threeparttable}
\end{center} 
\label{tab:ETA_comparision_results}
\end{table}

\section{Conclusion}
This paper proposed a privacy-preserving energy theft detection scheme joined with demand-response management for smart grid systems. The proposed scheme is the first to bridge the gap between the two issues. The accuracy of the proposed scheme's theft detection and demand estimation management are  analysed to confirm its  robustness against five types of energy theft attacks. In addition, two sets of privacy metrics are proposed and evaluated to ensure that the scheme ensures individual meter readings' privacy. The overall conclusion of all the experiments shows that the proposed scheme outperforms the existing privacy-preserving energy theft detectors in terms of detection rate and has significantly greater capabilities than other approaches as it can estimate the amount of theft along with the future demand with high accuracy.


%





\ifCLASSOPTIONcaptionsoff
  \newpage
\fi



%
\footnotesize
\bibliographystyle{IEEEtranN}
\bibliography{bare_jrnl}

%








\vfill

\end{document}